\documentclass[aps,onecolumn, 11pt]{revtex4-2}
\usepackage{amsmath,amssymb,amsfonts,color,graphicx,tabularx}
\usepackage[normalem]{ulem}
\usepackage{setspace}
\usepackage[unicode=true,colorlinks=true]{hyperref}
\usepackage{bm}
\hypersetup{linkcolor=blue,citecolor=blue,urlcolor=blue}
\renewcommand{\figurename}{\textbf{Fig.}}

\begin{document}

\title{Transdimensional anomalous Hall effect in rhombohedral thin graphite}

\author{Qingxin Li$^{1}$$^{\dagger}$}
\author{Hua Fan$^{2}$$^{\dagger}$}
\author{Min Li$^{3}$$^{\dagger}$}
\author{Yinghai Xu$^{1}$}
\author{Junwei Song$^{1}$}
\author{Kenji Watanabe$^{4}$}
\author{Takashi Taniguchi$^{5}$}
\author{Hua Jiang$^{6}$}
\author{Xin-Cheng Xie$^{6,7}$}
\author{James C. Hone$^{8}$}
\author{Cory R. Dean$^{9}$}
\author{Yue Zhao$^{2\ast}$}
\author{Jianpeng Liu$^{3,10\ast}$}
\author{Lei Wang$^{1,11\ast}$}

\affiliation{$^{1}$National key Laboratory of Solid-State Microstructures, School of Physics, Nanjing University, Nanjing, 210093, China}
\affiliation{$^{2}$Department of Physics, State key laboratory of quantum functional materials, Guangdong Basic Research Center of Excellence for Quantum Science, Southern University of Science and Technology (SUSTech), Shenzhen 518055, China}
\affiliation{$^{3}$School of Physical Science and Technology, ShanghaiTech Laboratory for Topological Physics, ShanghaiTech University, Shanghai 201210, China}
\affiliation{$^{4}$Research Center for Electronic and Optical Materials, National Institute for Materials Science, 1-1 Namiki, Tsukuba 305-0044, Japan}
\affiliation{$^{5}$Research Center for Materials Nanoarchitectonics, National Institute for Materials Science, 1-1 Namiki, Tsukuba 305-0044, Japan}
\affiliation{$^{6}$Interdisciplinary Center for Theoretical Physics and Information Sciences (ICTPIS), Fudan University, Shanghai 200433, China}
\affiliation{$^{7}$International Center for Quantum Materials, School of Physics, Peking University, Beijing 100871, China}
\affiliation{$^{8}$Department of Mechanical Engineering, Columbia University, New York, NY 10027, USA}
\affiliation{$^{9}$Department of Physics, Columbia University, New York, NY 10027, USA}
\affiliation{$^{10}$Liaoning Academy of Materials, Shenyang 110167, China}
\affiliation{$^{11}$Jiangsu Physical Science Research Center, Nanjing 210093, China}
\affiliation{$^{\dagger}$These authors contributed equally to this work.}
\affiliation{$^{\ast}$Corresponding authors, Email: zhaoy@sustech.edu.cn; liujp@shanghaitech.edu.cn; leiwang@nju.edu.cn}

\maketitle

\textbf{Anomalous Hall effect (AHE), occurring in materials with broken time-reversal symmetry, epitomizes the intricate interplay between magnetic order and orbital motions of electrons\cite{hall1881ferromagnets,karplus1954PR,Niuqian2004PRL,RoMP2010}. In two dimensional (2D) systems, AHE is always coupled with out-of-plane orbital magnetization associated in-plane chiral orbital motions. In three dimensional (3D) systems, carriers can tunnel or scatter along the third dimension within the vertical mean free path $\boldsymbol{l_z}$. When sample thickness far exceeds $\boldsymbol{l_z}$, scattering disrupts coherent out-of-plane motion, making 3D AHE effectively a thickness-averaged 2D counterpart\cite{RoMP2010}\textemdash still governed by out-of-plane orbital magnetization arising from in-plane orbital motions. Here, we explore an uncharted regime where the sample thickness is much larger than the atomic layer thickness yet smaller than or comparable to $\boldsymbol{l_z}$. In such ``transdimensional" regime, carriers can sustain coherent orbital motions both within and out of the 2D plane, leading to a fundamentally new type of AHE that couples both out-of-plane and in-plane orbital magnetizations. We report the first observation of such phenomenon\textemdash transdimensional AHE (TDAHE)\textemdash in electrostatically gated rhombohedral ennealayer graphene. This state emerges from a peculiar metallic phase that spontaneously breaks time-reversal, mirror and rotational symmetries driven by electron-electron interactions. Such TDAHE manifests as concurrent out-of-plane and in-plane Hall resistance hysteresis, controlled by external magnetic fields along either direction. Our findings unveils a new class of AHE, opening an unexplored paradigm for correlated and topological physics in transdimensional systems.}

\vspace{1em} \noindent\textbf{Introduction} \vspace{1em}
  
Electrons traveling through a material in the presence of a magnetic field follow a curved trajectory due to the Lorentz force, generating a Hall voltage perpendicular to both the current flow and the magnetic field directions\cite{hall1879}. While the ordinary Hall effect relies on an external magnetic field, the anomalous Hall effect (AHE) arises even without one, typically occurring in ferromagnetic systems with broken time-reversal symmetry\cite{RoMP2010}. AHE can arise from extrinsic mechanisms such as skew scattering\cite{SMIT1955skewAHE} and side jumps\cite{PhysRevB1970sideAHE}. More intriguingly, it can also originate from an intrinsic mechanism driven by momentum-space Berry curvature, which is fundamentally linked to the topological properties of conducting electrons\cite{RoMP2010,Niuqian2004PRL,karplus1954PR}. In conventional magnetic systems, the magnetization is usually contributed by both spin degrees of freedom and orbital magnetization. The latter results from the interaction between spin and the orbital motion of electrons via spin-orbit coupling (SOC), which is indispensable for the intrinsic AHE. Such a scenario applies to ferromagnetic metals~\cite{Niuqian2004PRL,karplus1954PR}, ferromagnetic semiconductors~\cite{MacDonald2002PRL} as well as magnetic topological materials~\cite{burkov2014PRL,ghimire2018NC,xue2013ScienceQAHE}. Defying this SOC-driven scenario, twisted van der Waals moir\'e systems have recently unveiled a different cause of AHE: it can emerge in a Coulomb-interaction-driven orbital ferromagnetic state which spontaneously breaks orbital time-reversal symmetry devoid of SOC~\cite{liu2021NRPorbital}. In such SOC-negligible system, both integer and fractional quantum anomalous Hall effects (QAHE) are observed~\cite{David2019Science,AFYoung2020scienceIntrinsic,julong2025NatureExtended,Xiaodong2023NatureFQAH,li2021NatureQAH}. Nevertheless, regardless of whether orbital magnetization stems from SOC or Coulomb interactions, all observed AHE and QAHE to date obey a fundamental orthogonality rule: the directions of orbital magnetization $\mathbf{M}$, current flow $\mathbf{J}$ and transverse electric field $\mathbf{E_{H}}$ associated with Hall voltage must be mutually perpendicular, i.e. $\mathbf{E_{\rm{H}}} \propto \mathbf{J} \times \mathbf{M}$.

\vspace{1em} \noindent\textbf{Dimensionality effect on AHE} \vspace{1em}

\begin{figure*}[t!]
\centering
\includegraphics[width=1\linewidth]{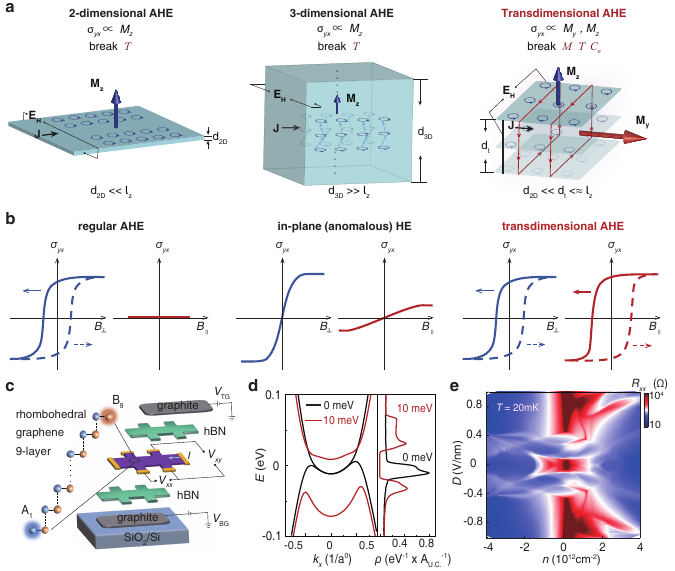}
\caption{
\textbf{Dimensionality perspective of AHE and basic characterizations on rhombohedral ennealayer graphene.} 
\textbf{a}, Illustration of the 2-dimensional AHE (left panel), 3-dimensional AHE (middle panel) and transdimensional AHE (right panel). The 2-dimensional AHE with broken time-reversal (\texorpdfstring{$\mathcal{T}$}{T}) symmetry is always coupled with out-of-plane orbital magnetization associated with in-plane chiral orbital motions. In 3-dimensional systems, carriers can tunnel or scatter along the third dimension within their vertical mean free path. When sample thickness is much greater than mean free path, scattering disrupts coherent out-of-plane motion, making the 3-dimensional AHE behave as the thickness-averaged 2D counterpart. The transdimensional AHE with broken \texorpdfstring{$\mathcal{T}$}{T}, n-fold rotational \texorpdfstring{$C_n$}{Cn}, and vertical mirror \texorpdfstring{$M_y$}{My} symmetries can host both out-of-plane magnetization and in-plane orbital magnetization.     
\textbf{b}, Illustration of the expected magnetic response of three different types of AHE. Regular AHE (left panel) with only out-of-plane magnetic hysteresis; in-plane (anomalous)HE (middle panel) with linear Hall resistance driven by in-plane external magnetic field; transdimensional AHE (right panel) with both out-of-plane and in-plane magnetic hysteresis. 
\textbf{c}, Schematic of the dual-gated Hall-bar device. Left: atomic structure of rhombohedral ennealayer graphene, with highlighted orbitals located at the bottom (\texorpdfstring{$A_1$}{A1}) and top (\texorpdfstring{$B_9$}{B9}) layers of ennealayer graphene dominating the wavefunctions of the low-energy bands.
\textbf{d}, Left panel: tight-binding calculations of the energy dispersion near the \texorpdfstring{\textbf{K}}{K} point of rhombohedral-stacked ennealayer graphene under an electric potential \texorpdfstring{$\Delta$}{Δ} = 0 meV (black) and \texorpdfstring{$\Delta$}{Δ} = 10 meV (red). Right panel: the single-particle density of states, \texorpdfstring{$\rho$}{ρ}, versus energy under \texorpdfstring{$\Delta$}{Δ} = 0 meV (black) and \texorpdfstring{$\Delta$}{Δ} = 10 meV (red).
\textbf{e}, Measured longitudinal resistance \texorpdfstring{$R_{xx}$}{Rxx} as a function of \texorpdfstring{$n$}{n} and \texorpdfstring{$D$}{D} at \texorpdfstring{$T$}{T} = 20 mK, and \texorpdfstring{$B$}{B} = 0 T. The corresponding \texorpdfstring{$\rho_{xy}$}{ρxy} map is shown in Extended Data Fig 3.
}
\label{fig:fig1}
\end{figure*}

A distinct yet long-overlooked perspective to unravel and classify AHE is the dimensionality in which it emerges.
In 2D (Fig.~\ref{fig:fig1}a left), AHE is characterized by chiral orbital motions within the plane, and is coupled with out-of-plane orbital magnetization $M_z$ (Fig.~\ref{fig:fig1}b left). AHE driven by in-plane spin magnetization, parallel to current flow, has been theoretically proposed in literature~\cite{liu2013PRL,ren2016PRB,liu2018PRL} (Fig.~\ref{fig:fig1}b middle). These approaches require specially designed form of SOC which couples in-plane spin magnetization to out-of-plane orbital magnetization. Therefore, this scenario can still be classified as the AHE generated by out-of-plane orbital magnetization.
In 3D (Fig.~\ref{fig:fig1}a middle), carriers are allowed to tunnel and/or get scattered along the third spatial dimension within a characteristic mean free path $l_z$. In such systems, the thickness of the sample $d$ is much larger than $l_z$ along the vertical direction. As a result, the coherent out-of-plane orbital motions of electrons would be overwhelmed by multiple scattering events within the sample. Thus the AHE of the 3D system would be manifested as its 2D counterpart averaged over the sample thickness, which can be considered as coupling with out-of-plane orbital magnetization too. However, an unexplored regime is when $d$ is much larger than the atomic layer thickness yet smaller than $l_z$ (Fig.~\ref{fig:fig1}a, b right). In such ``transdimensional" regime, carriers may undergo coherent orbital motions both within and out of the 2D plane. Under certain symmetry-broken conditions, this can give rise to a fundamentally new type of AHE that is coupled with both out-of-plane orbital magnetization $M_z$ and in-plane orbital magnetization $M_y$, a component parallel to current flow $\mathbf{J}$, which does not rigorously follow the $\mathbf{E_{\rm{H}}} \propto \mathbf{J} \times \mathbf{M}$ orthogonality rule. This points towards an unprecedented type of AHE which we term as transdimensional anomalous Hall effect (TDAHE). 

Here, we report the first observation of TDAHE in rhombohedral ennealayer graphene. Unlike thinner rhombohedral graphene (RG) structures that host only in-plane orbital motions and form conventional out-of-plane magnetization\cite{Guorui2024Science,julong2024ScienceLarge,julong2023NatureMultiferroicity,julong2024NatureFQAH}, our ennealayer layer (9-layer) RG possesses a $d$ value falling providentially within the transdimensional range - much exceeding the atomic layer thickness while smaller than $l_z$. Remarkably, in a spin and valley polarized region near van Hove singularity (VHS) facilitating strong $e-e$ interactions, we detect robust TDAHE manifested by coexisting out-of-plane and in-plane Hall resistance magnetic  hysteresis. These unusual hysteresis loop in both axes exhibit a striking magnetization dichotomy, small perpendicular versus large in-plane components, which can be controlled by external magnetic fields along either direction. 
Hartree-Fock calculations reveal that in the small doping and large $D$ regime, electron-electron interactions induce a spontaneous breaking of $C_n$ and $M_y$ mirror symmetries (thereby breaking $C_n$ and all mirror symmetries, as $M_x$ mirror symmetry is absent in rhombohedrally stacked graphene due to its stacking structure), which resulting in a Fermi surface with a crescent-shaped contour. This emergent asymmetric Fermi surface, combined with spontaneous time-reversal symmetry breaking, supports an in-plane magnetization parallel to the current flow for the first time, thereby endows a fundamentally new type of anomalous Hall effect - TDAHE.

\vspace{1em} \noindent\textbf{Phase diagram of rhombohedral ennealayer graphene} \vspace{1em}

The structure of our 9-layer RG device is shown in Fig.~\ref{fig:fig1}c, which contains top and bottom graphite gates with voltages $V_{TG}$ and $V_{BG}$, enabling us to independently tune carrier density $n$ and  $D$. The low-energy sublattice sites localized in the outermost layer (marked as two highlighted sublattices) causes $E \propto \pm \lvert k \rvert ^N$ energy dispersion, making the $D$-tuned DOS of 9-layer RG is significantly large (Fig.~\ref{fig:fig1}d). Fig.~\ref{fig:fig1}e shows the longitudinal resistance $R_{xx}$ versus $n$ and $D$ at magnetic field  $B$ = 0 and $T$ = 15 mK. At $n$ = 0, we observe an insulating phases at high $\lvert D \rvert$ associated with the layer-polarized state and another insulator at $D$ = 0 which defies the single-particle picture (Fig.~\ref{fig:fig1}d) (we called correlated insulator (CI) in the following text). The temperature-dependent flattening of CI (Extended Data Fig 2) indicates a spontaneous symmetry-breaking transition.  

\begin{figure*}[t!]
\begin{center}
\includegraphics[width=1\linewidth]{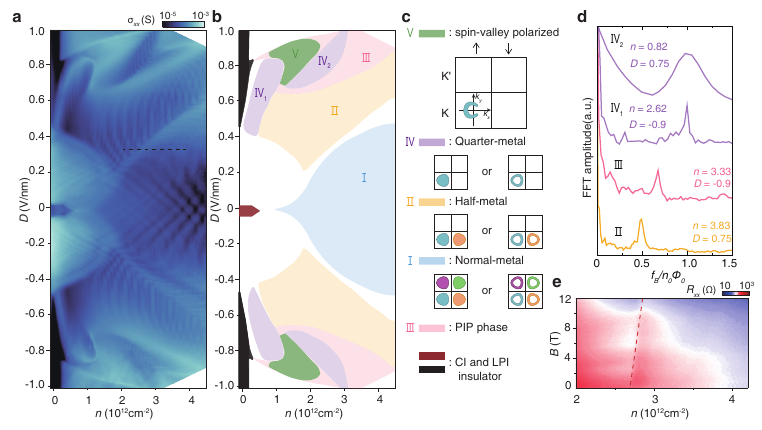}
\caption
{
\textbf{Phase diagram and fermiology in the conduction band.} 
    \textbf{a}, Map of the longitudinal conductivity $\sigma_{xx}$ plotted versus $n$ and the displacement field $D$ under a magnetic field $B_\perp=4$ T and base temperature $T$ = 15 mK. 
    \textbf{b}, Schematic diagram illustrating the main phases shown in \textbf{a}. The differently coloured regions, labeled with distinct Roman numerals, represent distinct spin- and valley-polarized flavours: \uppercase\expandafter{\romannumeral1} normal metal, \uppercase\expandafter{\romannumeral2} half metal, \uppercase\expandafter{\romannumeral3} partial isospin polarized metal and \uppercase\expandafter{\romannumeral4} quarter metal. In here, we find an unreported  metallic phase (green region) in \uppercase\expandafter{\romannumeral5}. This state persists without any quantum oscillations and Landau levels formation up to $B_\perp$ = 13 T (see Extended Data Fig 5). The TDAHE state is in this region. 
    \textbf{c}, Illustration of various symmetry-broken metallic phases and the CI, layer-polarized insulator (LPI) in \textbf{b}, along with schematically possible Fermi surface contours extracted based on Fourier transforms. The filled colours in Fermi surface contours represent the spin and valley flavours. The phase \uppercase\expandafter{\romannumeral5} exhibits a new peculiar fermi surface with crescent shape, breaking $C_n$ and $M_y$ symmetries which is not observed in previous studies (see more details in Fig. 4d and main-text). 
     \textbf{d}, Fast Fourier transformation (FFT) analysis of the quantum oscillations in QM, PIP and HM phases. The specified values of $n$ and $D$ are expressed in units of $10^{12}$ $\mathrm{cm}^{\mathrm{-2}}$ and V/nm.
     \textbf{e}, Map of $R_{xx}$ as a function of $n$ and $B_{\parallel}$ at $D$ = 0.35 V/nm. The data are taken along the horizontal black dashed line in \textbf{a}. The red dashed line corresponds to the boundary of HM phase with spin-polarization and NM phase without spin-polarization.
}
\label{fig:fig2}
\end{center}
\end{figure*}

 Beyond the spontaneous symmetry-breaking insulator at charge neutrality (CNP), we also uncover spontaneously symmetry-broken metals in small $n$ and large $D$ regimes, where $D$-enhanced VHSs lead to divergent DOS for the emergence of strongly correlated electronic states. Fig.~\ref{fig:fig2}a shows $\sigma_{xx}$ vs $n$ and $D$ at $B_\perp$ = 4 T on electron-doped side. A series of transitions manifesting as the white-line feature divide the metallic phase into several regions with distinct spin and valley flavour polarization~\cite{stoner1938collective,zhou2021NatureHalf,zhou2022ScienceIsospin} (schematically shown in Fig.~\ref{fig:fig2}b). The phase \uppercase\expandafter{\romannumeral1} harbours LLs with fourfold degeneracy, which corresponds to normal metal (NM) combined spin and valley degeneracies; the phase \uppercase\expandafter{\romannumeral2} and \uppercase\expandafter{\romannumeral4}, where the degeneracies are reduced to twofold and onefold (identify by Fourier transforms in Fig.~\ref{fig:fig2}d), correspond to a quarter-metal (QM) and half-metal (HM), respectively. Notably, in phase \uppercase\expandafter{\romannumeral4},  strong electronic interactions  render the system prone to emerge exchange-interaction-driven ferromagnetic ground states\cite{stoner1938collective,zhou2021NatureHalf,zhou2022ScienceIsospin}, exhibiting Stoner ferromagnetism with out-of-plane magnetic hysteresis (Extended Data Fig 4). Besides, the evolution of phase \uppercase\expandafter{\romannumeral3} with the $B_{\parallel}$ (Fig.~\ref{fig:fig2}f) suggests a spin-polarized but valley-unpolarized phase that is stabilized by $B_{\parallel}$. Finally, the phase \uppercase\expandafter{\romannumeral3} shows an oscillation frequency 0.5 $<$ $f_{\nu}$ $<$ 1 (Fig.~\ref{fig:fig2}e), corresponding to the partially isospin polarized (PIP) phase where the single Fermi surface simultaneously accommodates two majority and two minority isospin flavors.

\vspace{1em} \noindent\textbf{Transdimensional anomalous Hall state} \vspace{1em}

In the spontaneously symmetry-broken metallic phases, phase \uppercase\expandafter{\romannumeral5} (green region) exhibits a striking absence of quantum Shubnikov-de Haas (SdH) oscillations with increasing  $B_\perp$. This phase is situated between two QM phases and a PIP phase in the $\sigma_{xx}$($n$, $D$) map, marked by elevated resistance in both $R_{xx}$ and$R_{xy}$ compared to neighboring phases (Fig. 3a, Extended Data Fig 3). Surprisingly, phase \uppercase\expandafter{\romannumeral5} persists without LL formation up to $B_\perp$ = 13 T (Extended Data Fig 5), indicating a robust strongly correlated ground state that emerges from enhanced $e-e$ correlations at low density and large $D$. Under strong magnetic field, this correlation-driven state prevails over quantum Hall states, suppressing the conventional LL features.

\begin{figure*}[t!]
\begin{center}
\includegraphics[width=1\linewidth]{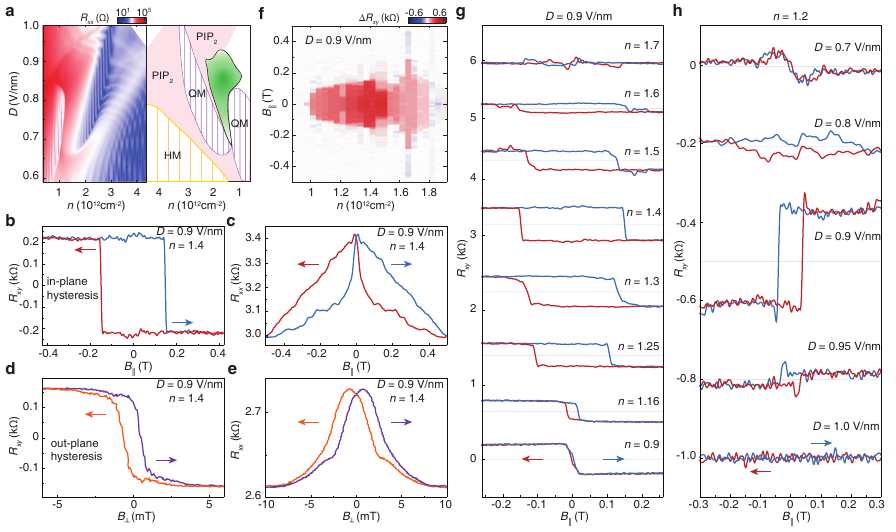}
\caption
{
\textbf{TDAHE with unique in-plane orbital magnetization  and out-of-plane magnetization.} 
    \textbf{a}, Map of $R_{xx}$ plotted versus $D$ and $n$ at $B_\perp$ = 4 T, for $0.6$ V/nm ${\leqslant}D{\leqslant}1.0$ V/nm. Illustration of different isospin flavours and TDAHE state (green region) are depicted in right panel.
    \textbf{b-c}, Antisymmetrized Hall resistance $R_{xy}$ (\textbf{b}) and symmetrized  $R_{xx}$ (\textbf{c}) measured as $B_{\parallel}$ is swept back and forth (marked by red and blue arrows) at $n$ = 1.4 x $10^{12}$ $\mathrm{cm}^{\mathrm{-2}}$ and $D$ = 0.9 V/nm. The pronounced hysteresis loop with $B_{\parallel}$ indicates the presence of in-plane orbital ferromagnetism. 
    \textbf{d-e}, $R_{xy}$ (\textbf{d}) and $R_{xx}$ (\textbf{e}) measured as $B_\perp$ is swept back and forth (marked by orange and purple arrows) at $n$ = 1.4 x $10^{12}$ $\mathrm{cm}^{\mathrm{-2}}$ and $D$ = 0.9 V/nm. The coercive field, comparable to that observed in the QM phase, indicates a Stoner-type isospin ferromagnetic contribution for magnetization.  
    \textbf{f},  Hall resistance difference $\Delta R_{xy}$ ($\Delta R_{xy}$ = $R_{xy}^{B_{\parallel} \uparrow}$ - $R_{xy}^{B_{\parallel} \downarrow}$) as a function of $n$ and $B_{\parallel}$ at $D$ = 0.9 V/nm. 
    \textbf{g}, The parallel magnetic-field dependence of $R_{xy}$ at several selected carrier density $n$ values (in units of $10^{12}$ $\mathrm{cm}^{\mathrm{-2}}$) at $D$ = 0.9 V/nm , T = 50 mK. 
    \textbf{h}, The parallel-magnetic-field dependence of $R_{xy}$ at several selected displacement fields $D$ values at $n$ = 1.2 x $10^{12}$ $\mathrm{cm}^{\mathrm{-2}}$ and T = 50 mK.
}
\label{fig:fig3}
\end{center}
\end{figure*}

In order to determine the ground state of this phase, we investigate the magnetic character by measuring the magnetic field dependence of the $R_{xx}$ and Hall resistance $R_{xy}$ for both $B_{\parallel}$ and $B_\perp$. Amazingly, we observe a magnetic hysteresis loop with $B_\perp$, and an unexpected pronounced hysteresis loop with $B_{\parallel}$, as shown in Fig.~\ref{fig:fig3}b and Fig.~\ref{fig:fig3}d. The $B_\perp$-induced $R_{xy}$ hysteresis loop, beginning at approximately $B_\perp$ = $\pm$3 mT, shows a coercive field comparable to that observed in the QM phase (Extended Data Fig 4). Thus, this hysteresis may primarily originate from the Stoner-type isospin ferromagnetism. However, the coercive field of $B_{\parallel}$-induced $R_{xy}$ hysteresis loop is about 160mT (can reach up to as large as 500 mT under different $n$ and $D$), which is significantly larger than the contribution from isospin orders. Meanwhile, as shown in Fig.~\ref{fig:fig3}c and Fig.~\ref{fig:fig3}e, the $B_{\parallel}$-induced and $B_\perp$-induced hysteresis loop also appear in $R_{xx}$ measurement. The unambiguous magnetic hysteresis loop observed in 9-layer RG indicates the presence of TDAHE with coexisting the unique in-plane orbital magnetization and the out-of-plane magnetization. Unlike the established paradigm of exchange-interaction-driven Stoner ferromagnetism, the giant in-plane orbital ferromagnetism discovered here represents the first observation of its kind. We now turn to a detailed investigation of this in-plane orbital ferromagnetism induced AHE.

\vspace{1em} \noindent\textbf{Doping and displacement field tuning} \vspace{1em}

We next investigate the $n$-dependence and the $D$-dependence of the in-plane orbital ferromagnetism. Fig.~\ref{fig:fig3}h illustrates the $D$ range where in-plane magnetization occurs. With $n$ fixed at 1.2 x $10^{12}$ $\mathrm{cm}^{\mathrm{-2}}$, accompanying the disappearance of the QM phase, the in-plane orbital magnetization emerges at $D$ = 0.7 V/nm. As $D$ increases, the magnitude of AHE  initially enhances and then diminishes, exhibiting a maximum anomalous Hall signal at $D$ = 0.9 V/nm, eventually vanishes at D = 1.0 V/nm, where the PIP phase begins to develop. Fig.~\ref{fig:fig3}g  shows a non-monotonic variation of the in-plane orbital magnetization  with $n$ at $D$ = 0.9 V/nm. The $B_{\parallel}$-induced hysteresis loop appears at $n$ = 1 x $10^{12}$ $\mathrm{cm}^{\mathrm{-2}}$, and with further increases in density, both the Hall resistance and coercive field of  hysteresis loop increase. The Hall resistance reaches its maximum at $n$ = 1.4 x $10^{12}$ $\mathrm{cm}^{\mathrm{-2}}$. Beyond this point, the AHE signal gradually decreases and collapses at $n$ = 2 × $10^{12}$ $\mathrm{cm}^{\mathrm{-2}}$, where the QM phase starts to dominate. Notably, near the QM - TDAHE phase boundary, we observe electric field  hysteresis loop shown in Extended Data Fig 7. This hysteretic behavior may signal a discontinuous first-order phase transition separating two distinct Fermi surface topologies\cite{zhou2021NatureHalf}, indicative of competing ground states at the QM - TDAHE interface. 

Fig.~\ref{fig:fig3}f maps the Hall resistance difference $\Delta R_{xy}$ between forward and backward $B$ sweeps ($\Delta R_{xy}$ = $R_{xy}^{B_{\parallel} \uparrow}$ - $R_{xy}^{B_{\parallel} \downarrow}$) as a function of $n$ and $B_{\parallel}$ at $D$ = 0.9 V/nm. From the picture, we observe not only the overall evolution of the in-plane magnetization, from its emergence to strengthening, weakening, and eventual collapsing, but also evidence of the switching magnetic order behavior occurring near the density where the  hysteresis loop almost vanishes (marked by the red arrow in Extended Data Fig 6). Doping induced switching magnetic order in Chern insulators has been reported previously in twisted and rhombohedral graphene multilayers systems~\cite{AFYoung2020NatureElectrical,Guorui2024Science,choi2025superconductivity}, where the magnetization changes sign within a phase with fixed valley polarization~\cite{MacDonald2020PRLvoltage}. In the case of orbital Chern insulator in moiré system with  large unit cell areas, the topological edge states can induce a giant magnetization jump, which has the opposite sign compared to the total magnetization,  driving a sign change in the overall magnetization~\cite{MacDonald2020PRLvoltage,AFYoung2020NatureElectrical}. However, in our 9-layer RG without moir\'e superlattices potential, the magnetization reversal appears in a metallic phase without the topological gap, which is unlike the case in Chern insulators. We believe that the magnetization reversal is more likely to arise from the density-dependent orbital magnetization of Bloch states with finite Berry curvature\cite{choi2025superconductivity}.

\vspace{1em} \noindent\textbf{Temperature dependence} \vspace{1em}

\begin{figure*}[t!]
\begin{center}
\includegraphics[width=1\linewidth]{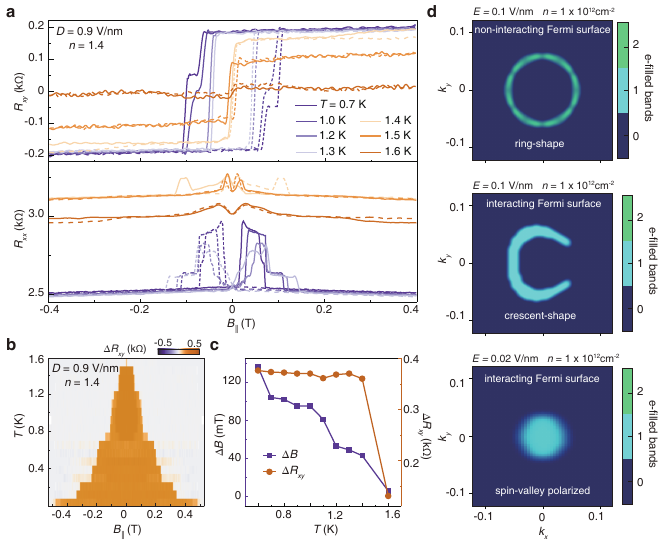}
\caption
{
\textbf{Temperature dependence of in-plane magnetization and Hartree-Fock calculations.} 
    \textbf{a}, Magnetic field dependence of $R_{xy}$ and $R_{xx}$ at selected temperatures 
    with a displacement field of $D$ = 0.9 V/nm and a carrier density $n$ = 1.4 x $10^{12}$ $\mathrm{cm}^{\mathrm{-2}}$.
    \textbf{b}, Hall resistance difference $\Delta R_{xy}$ as a function of $T$ and $B_{\parallel}$ at $n$ = 1.4 x $10^{12}$ $\mathrm{cm}^{\mathrm{-2}}$ and $D$ = 0.9 V/nm. The coercive field of $B_{\parallel}$-induced $R_{xy}$ hysteresis loop can reach up to as large as 500 mT at 20 mK. The magnitude of the in-plane orbital ferromagnetism is suppressed with increasing temperature, evidenced by shrinking coercive field and lowering the $\Delta R_{xy}$. 
    \textbf{c}, Coercive field $\Delta B$ and $\Delta R_{xy}$ plotted as a function of $T$ from the same data partially shown in \textbf{b}.  With increasing $T$, the $\Delta R_{xy}$ signal remaining nearly constant up to 1.5 K and then dropping sharply above 1.6 K; The Coercive field continuously decreases and vanishes at 1.6 K.
    \textbf{d}, Hartree-Fock calculations of Fermi contours. The different colours in the colourbar represent different numbers of isospins in the electron-filled energy bands. The upper panel plots the ring-shaped non-interacting Fermi surface of the system when vertical electric field $E \gtrapprox 0.03\rm{-}0.05$\,V/nm. The middle panel depicts  a peculiar metallic state with a crescent-shaped Fermi surface incorporating dominant long-range Coulomb electron-electron interactions, exhibiting complete spin and valley polarization. In this state, the system spontaneously breaks time-reversal ($\mathcal{T}$), three-fold rotational ($C_3$), and vertical mirror ($M_y$) symmetries. This symmetry breaking generates a substantial in-plane orbital magnetization, while producing non-zero anomalous Hall conductivity through $\mathcal{T}$-symmetry violation. We call such state as ``transdimensional orbital ferromagnetic state". The lower panel shows a regular interacting spin-valley polarized Fermi surface when electric field is small (or when the carrier density is large), which preserves $C_3$ symmetry thus kills in-plane orbital magnetization. 
}
\label{fig:fig4}
\end{center}
\end{figure*}

With $n$ fixed at 1.4 x $10^{12}$ $\mathrm{cm}^{\mathrm{-2}}$ and $D$ = 0.9 V/nm, where the in-plane orbital magnetization reaches its maximum value, we examine its temperature dependence.  Fig.~\ref{fig:fig4}a and Fig.~\ref{fig:fig4}b show the temperature ($T$) dependence of $B_{\parallel}$-induced $R_{xy}$ and $R_{xx}$ hysteresis loop and Fig.~\ref{fig:fig4}c plots the map of $\Delta R_{xy}$ as function of $B_{\parallel}$ and $T$. The magnitude of the in-plane orbital ferromagnetism  is suppressed with increasing temperature, evidenced by shrinking coercive field and lowering the $\Delta R_{xy}$.  Remarkably, the $\Delta R_{xy}$ signal remains large at temperatures near 1.5 K, indicating the relatively large energy scales associated with in-plane orbital magnetism in 9-layer RG. With increasing $T$, the coercive field continuously decreases and vanishes at 1.6 K (Fig.~\ref{fig:fig4}c), while the anomalous Hall resistance magnitude exhibits a distinct temperature dependence, remaining nearly constant up to 1.5 K and then dropping sharply above 1.6 K.

The suppression of in-plane orbital ferromagnetism with increasing temperature may result from two possible mechanisms. The first involves thermal activation, which can promote the flipping of individual magnetic domains or the motion of domain walls, leading to a monotonic temperature dependence of the coercive field~\cite{emori2015generalized, David2019Science}. The second concerns a temperature-induced reduction in mean free path $l_z$. When the $l_z$ becomes much shorter than the thickness of the device, the coherent out-of-plane orbital motions vanish, leading to the collapse of in-plane orbital magnetization.

\vspace{1em} \noindent\textbf{Discussion} \vspace{1em}

Our observation of the giant in-plane orbital magnetization is pretty unexpected. While previous theoretical studies have explored potential mechanisms for realizing in-plane magnetization AHE or QAHE~\cite{liu2013PRL,liu2018PRL,ren2016PRB,CaoPRLipAHE} in strongly spin-orbit coupled topological materials, the experimental realization of these effects remains highly challenging. The emergence of in-plane magnetization AHE or QAHE requires not only the breaking of time-reversal symmetry through magnetization along an arbitrary direction but also other stringent symmetry conditions, such as breaking all reflection symmetries~\cite{liu2013PRL}. However, the intrinsic symmetry constraints inherent in most material systems have posed significant barriers to achieving these effects in experimental settings. Recent experimental studies have reported the observation of an in-plane Hall effect under zero $B_\perp$ in two distinct material systems: the antiferromagnetic heterodimensional V$\mathrm{S}_{\mathrm{2}}$-VS superlattice and the magnetic Weyl semimetal EuC$\mathrm{d}_{\mathrm{2}}$S$\mathrm{d}_{\mathrm{2}}$~~\cite{nakamura2024PRL,zhou2022NatureIPHE}. In those works, it is noteworthy that those observations were made under a non-zero $B_{\parallel}$, and unfortunately, no anomalous Hall signal was detected for zero value of $B_{\parallel}$ and $B_\perp$. In graphene-based systems, achieving in-plane magnetic states with  simultaneous breaking both time-reversal symmetry and all reflection symmetries is even more exceptionally challenging due to the high symmetry of lattice structure, the absence of transition metal  elements and the negligibly weak SOC ($\sim$ 40 $\mathrm{\mu}$eV)~\cite{AFYoung2024NpIntervalley,kane2005QSH} in the graphite.

To figure out the origin of in-plane orbital magnetization, we conduct systematic unrestricted Hartree-Fock calculations on the 9-layer rhombohedral graphene (RG) system (see the supplementary file). We find that as long as the vertical electric field $E \gtrapprox 0.03\rm{-}0.05$\,V/nm, the non-interacting Fermi surface of the system upon slight electron doping $n\lessapprox 10^{12}\,$cm$^{-2}$ forms a ring-shape as shown in the upper panel of Fig.~4(d). Including the dominant long-range electron-electron Coulomb interactions would drive the system to  a peculiar metallic state which spontaneously break time-reversal ($\mathcal{T}$), three-fold rotation ($C_3$) and vertical mirror ($M_y$) symmetries of the system. This metallic phase is characterized by a crescent-shaped Fermi surface with full spin and valley polarization, as shown in the middle panel of Fig.~4(d). It possesses a notable in-plane orbital magnetization on the order of $2.2\,\mu_B$ per electron thanks to the breaking of $\mathcal{T}$, $C_3$ and $M_y$ symmetries, and exhibits non-vanishing anomalous Hall conductivity due to the breaking of $\mathcal{T}$ symmetry. By virtue of non-vanishing in-plane orbital magnetization, this state naturally couples to in-plane magnetic field and its anomalous Hall conductivity flips sign with the reversal of in-plane field direction.  We call such state as ``transdimensional orbital ferromagnetic state". More detailed mechanism and properties of transdimensional orbital magnetism will be reported in a separate theoretical work\cite{theoryTDAHE}. When electric field is small (or when the carrier density is large), the interacting ground state of electron-doped 9-layer rhombohedral graphene becomes a regular spin-valley polarized metal whose Fermi surface is shown in the lower panel of Fig.~4(d), which preserves $C_3$ symmetry thus kills in-plane orbital magnetization.

It is worth noting that the in-plane magnetization in our case is of pure orbital nature, the emergence of which is solely driven by electron-electron interactions without the need for SOC. Electrons in such metallic ground state would form an out-of-plane current loop as schematically shown in the right panel of  Fig.~1(a), thus generating in-plane orbital magnetization. In contrast, in all of the previous theoretical proposals of AHE induced by in-plane magnetization, it is the spin magnetization that is within the 2D plane\cite{liu2013PRL,liu2018PRL,ren2016PRB,CaoPRLipAHE}; some specifically designed form of SOC is needed to convert the in-plane spin magnetization to out-of-plane orbital magnetization in order to generate AHE. In this sense, the TDAHE reported in this work is a genuine new class of AHE, and is fundamentally different from the previous theoretical proposals of in-plane-magnetization induced AHE. This establishes a new paradigm for correlation-driven topological phases beyond conventional AHE classifications. Moving forward, more exotic states such as integer- or fractional- quantized TDAHE states protected by topologically nontrivial Chern gaps may be realized by introducing tailored moir\'e potentials. These would require future explorations of  heterostructure and supercell engineerings in such transdimensional regime.

\vspace{1em} \noindent\textbf{Methods} \vspace{1em}

The Rhombohedral ennealayer graphene and hBN flakes are prepared by mechanical exfoliation of bulk crystals. The rhombohedral domains of ennealayer graphene flakes are detected by a confocal Raman system (WITec Alpha 300R) under a 532 nm excitation laser at room temperature. The rhombohedral domains are subsequently isolated using a anodic-oxidation-assisted atomic force microscope cutting~\cite{AFM2009APL,AFM2018Nanoletter}. The Van der Waals heterostructures are fabricated using our ``pick-up method"~\cite{wang2013one} to achieve a multi-layer heterostructure with the  ennealayer graphene encapsulated by two flakes of hexagonal boron nitride (hBN) and thin graphite flakes as the top and bottom gates. Electron-beam lithography is used to write an etch mask to define the Hall-bar geometry and the electrodes. Redundant regions are etched away by CHF$_{3}$/O$_{2}$ plasma. Finally the ennealayer graphene and gates are edge-contacted by e-beam evaporating thin metal layers consisting of Cr/Pd/Au (1 nm/15 nm/100 nm). 
 
The transport measurements are performed in two systems, a dilution fridge with a base temperature of 15 mK and a VTI fridge down to 1.5 K, and both are with superconducting magnets. All data are taken using the standard four-terminal configuration with lock-in amplifier techniques by sourcing an AC current $I$ between 2 and 10 nA at a frequency of 17.777 Hz. Top and bottom graphite gates with voltages $V_{T}$ and $V_{B}$, allows us to independently tune carrier density: $n=(C_{B}V_{B}+C_{T}V_{T})/e$, and displacement field: $D=(C_{B}V_{B}-C_{T}V_{T})/2$, where $C_{T}$($C_{B}$) is top (bottom) gate capacitance. The data of longitudinal conductivity $\sigma_{xx}$ and Hall conductivity $\sigma_{xy}$ are obtained from the measured resistances by $\sigma_{xx}=\rho_{xx}/(\rho_{xx}^2+R_{xy}^2)$. The data presented in the main text is based on device A, and additional data from device B are included in Extended Data Fig 9.

We note that finite couplings between longitudinal and transverse directions would mix $R_{xx}$ and $R_{xy}$ signals. According to that $R_{xy}$ is antisymmetric with respect to magnetic field $B$, whereas $R_{xx}$ is symmetric, we can separate the two components using standard symmetrization and anti-symmetrization procedures during the data processing. The symmetrization procedure for $R_{xx}$ is as follows: $R_{xx}^{sym} (B, \uparrow)$ = $((R_{xx} (B, \uparrow) + R_{xx} (-B, \downarrow))/2$, and $R_{xx}^{sym} (B, \downarrow)$ = $((R_{xx} (B, \downarrow) + R_{xx} (-B, \uparrow))/2$. The anti-symmetrization procedure for $R_{xy}$ is as follows: $R_{xy}^{antisym} (B, \uparrow)$ = $((R_{xy} (B, \uparrow) - R_{xy} (-B, \downarrow))/2$, and $R_{xy}^{antisym} (B, \downarrow)$ = $((R_{xy} (B, \downarrow) - R_{xy} (-B, \uparrow))/2$. where $\uparrow/\downarrow$ indicates the direction of the magnetic field sweeping.

\section*{D\MakeLowercase{ata availability}}

The data that support the findings of this study are available from the corresponding authors upon request.

\section*{C\MakeLowercase{ode availability}}

The code that support the findings of this study is available from the corresponding authors upon request.

\section*{R\MakeLowercase{eferences}}
\bibliography{refs.bib}

\section*{A\MakeLowercase{cknowledgments}}

L.W. acknowledges support from the National Key Projects for Research and Development of China (Grant Nos. 2021YFA1400400, 2022YFA1204700) and Natural Science Foundation of Jiangsu Province (Grant Nos. BK20220066 and BK20233001). K.W. and T.T. acknowledge support from the JSPS KAKENHI (Grant Numbers 21H05233 and 23H02052) and World Premier International Research Center Initiative (WPI), MEXT, Japan. J.L. acknowledges support from the  National Key Research and Development Program of China (Grant No. 2024YFA1410400) and  the National Natural Science Foundation of China (Grant No. 12174257). Y.Z. acknowledges support from the National Key R$\&$D Program of China (Grants No. 2022YFA1403700), NSFC 11674150, University Innovative Team in Guangdong Province 2020KCXTD001.

\vspace{1em} \noindent\textbf{Author Contributions} \vspace{1em}

L.W. conceived and supervised the experiment. Q.L., Y.X. and J.S. fabricated the samples. Q.L., H.F., and Y.Z. performed the transport measurements. Q.L., L.W. Y.Z., H.J., X.C.X., J.C.H., C.R.D and J.L. analyzed the data. M.L. and J.L. conducted theoretical calculations. K.W. and T.T supplied the hBN crystals. Q.L., J.L. and L.W. wrote the manuscript with input from all co-authors.

\vspace{1em} \noindent\textbf{Competing interests} \vspace{1em}

The authors declare no competing interest.

\newpage
\clearpage

\renewcommand{\figurename}{\textbf{Extended Data Fig}}
\setcounter{figure}{0}
\setcounter{equation}{0}
\vspace{1em} \noindent\textbf{Extended Data Figures} \vspace{1em}

\begin{figure}[h!]
\begin{center}
\includegraphics[width=1\linewidth]{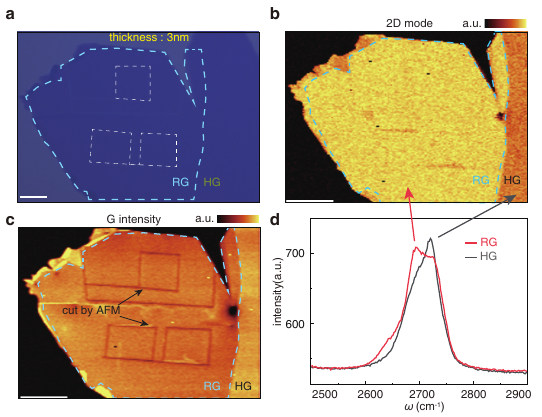}
\caption
{
\textbf{Identification of 9-layer RG.} 
       \textbf{a}, Optical image of an exfoliated 3 nm 9-layer graphene flake on SiO$_2$/Si substrate, taken with a regular optical microscope. The Cyan dashed line denotes the boundary of rhombohedral stacked (RG) region (ABCABC...) and the RG region is isolated by anodic-oxidation-assisted AFM cutting denoted by the white dashed lines. The scalebar corresponds to 10 $\mu$m.
       \textbf{b}, Raman spectrum map of the 9-layer graphene flake in panel \textbf{a}. The color represents the peak position of the 2D mode and the darker domain indicates the hexagonal-stacked (ABABAB...) 9-layer graphene, and the brighter domain is defined as rhombohedral stacking enclosed by Cyan dashed line. The scalebar corresponds to 10 $\mu$m.
       \textbf{c}, Raman mapping of the integrated G band intensity. The brighter domain indicates the hexagonal-stacked  9-layer graphene, and the darker domain is defined as rhombohedral stacking enclosed by Cyan dashed line.  The scalebar corresponds to 10 $\mu$m.
       \textbf{d}, Typical Raman spectra of hexagonal-stacked (HG) and RG of 9-layer graphene, centered on the 2D mode.
}
\label{fig:fig.e1}
\end{center}
\end{figure}

\newpage
\begin{figure*}[h!]
\includegraphics[width=0.8\linewidth]{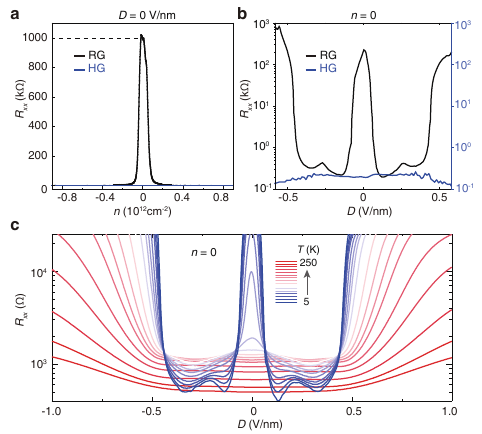}
\caption
{
\textbf{Temperature dependence of CI in $n=D=0$.}
      \textbf{a}, Longitudinal resistance $R_{xx}$ as a function of $n$ at $D$ = 0 V/nm, $T$ = 20 mK, $B$ = 0 T, with RG (black curve) and HG (blue curve) crystal structures. The black dashed line marks the resistance of 1000 k$\Omega $.
      \textbf{b}, $R_{xx}$ as a function of $D$ at $n$ = 0, $T$ = 20 mK, $B$ = 0 T, with RG (black curve) and HG (blue curve) crystal structures. 
      \textbf{c}, Temperature dependence of the $R_{xx}$ verse $D$ measured at $n=0$ from 5 K to 250 K.
}
\label{fig:fig.e2}
\end{figure*}

\newpage
\begin{figure*}[h!]
\includegraphics[width=0.8\linewidth]{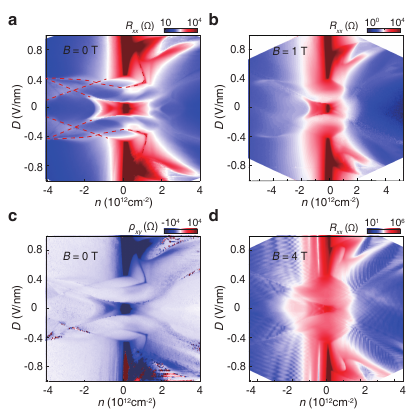}
\caption
{
\textbf{Phase diagram of 9-layer RG in different out-of-plane magnetic fields.} 
     \textbf{a}, Map of longitudinal resistance $R_{xx}$ as a function of $n$ and $D$ at $T$ = 20 mK, $B$ = 0 T. The red dashed lines denotes several cross-like features occurring with increasing resistance, which may be associated with van Hove singularities.
     \textbf{b}, Map of longitudinal resistance $R_{xx}$ as a function of $n$ and $D$ at $T$ = 20 mK, $B_\perp$ = 1 T.
      \textbf{c}, Map of Hall resistivity $\rho_{xy}$ as a function of $n$ and $D$ at $T$ = 20 mK, $B$ = 0 T.
      \textbf{d}, Map of longitudinal resistance $R_{xx}$ as a function of $n$ and $D$ at $T$ = 20 mK, $B_\perp$ = 4 T.
}
\label{fig:fig.e3}
\end{figure*}

\newpage
\begin{figure*}[h!]
\includegraphics[width=0.8\linewidth]{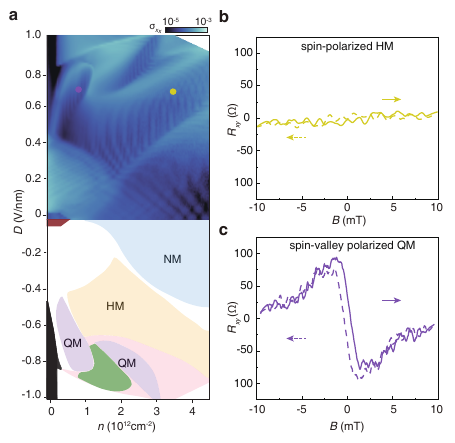}
\caption
{
\textbf{Stoner ferromagnetism in QM region.} 
     \textbf{a}, Longitudinal conductivity $\sigma_{xx}$ plotted versus $n$ and the displacement field $D$ for $0$ V/nm ${\leqslant}D{\leqslant}1.0$ V/nm, $B_\perp$ = 4 T, and the corresponding schematic diagram for different symmetry-broken phases.
     \textbf{b-c}, $R_{xy}$ measured as $B_\perp$ is swept back and forth (marked by dashed and solid arrows) at yellow (\textbf{b}) and purple (\textbf{c}) points marked in \textbf{a}.
}
\label{fig:fig.e4}
\end{figure*}

\newpage
\begin{figure*}[h!]
\includegraphics[width=0.9\linewidth]{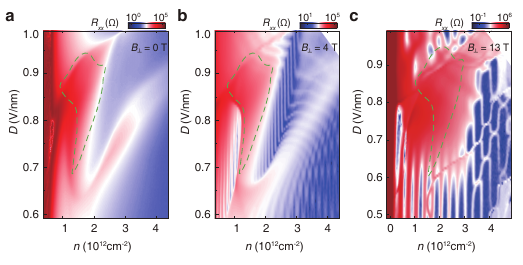}
\caption
{
\textbf{TDAHE phase under different out-of-plane magnetic fields: $B_\perp$ = 0 T, 4 T, 13 T.} 
     \textbf{a-c}, $R_{xx}$ plotted versus $n$ and $D$ for $0.6$ V/nm ${\leqslant}D{\leqslant}1.0$ V/nm, at $B$ = 0 T (\textbf{a}), $B_\perp$ = 4 T (\textbf{b}), $B_\perp$ = 13 T (\textbf{c}). The green dashed line in each figure denotes the boundary of TDAHE region.
}
\label{fig:fig.e5}
\end{figure*}

\newpage
\begin{figure*}[h!]
\includegraphics[width=0.6\linewidth]{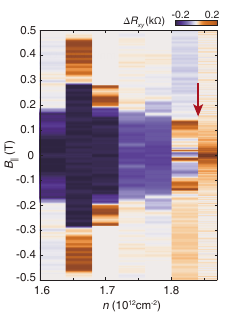}
\caption
{
\textbf{Electrical switching of the magnetic order.} 
     Hall resistance difference $\Delta R_{xy}$ as a function of $n$ and $B_{\parallel}$ at $D$ = 0.9 V/nm, for $1.6\times10^{12}$ $\mathrm{cm}^{\mathrm{-2}}$ ${\leqslant}n{\leqslant}1.9\times10^{12}$ $\mathrm{cm}^{\mathrm{-2}}$. The switching magnetic order behavior occurs near the density where the  hystereses loop almost vanishes (marked by the red arrow). 
}
\label{fig:fig.e6}
\end{figure*}

\newpage
\begin{figure*}[h!]
\includegraphics[width=1.0\linewidth]{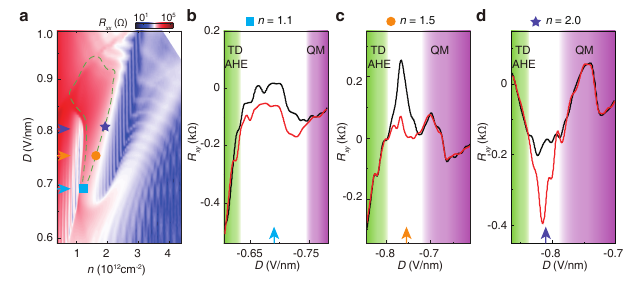}
\caption
{
\textbf{Electrical hysteresis near the boundary of TDAHE and QM.} 
     \textbf{a}, $R_{xx}$ plotted versus $n$ and $D$, at $B_\perp$ = 4 T. The green dashed line denotes the boundary of TDAHE region.
     \textbf{b-d}, Hall resistance $R_{xy}$ measured as a function of $D$ across the boundary between the QM and TDAHE phases, at the locations marked by the blue square (\textbf{b}), orange sphere (\textbf{c}), and purple pentagon (\textbf{d}) in \textbf{a}. The transition is strongly first order, showing hysteresis as a function of applied gate voltages.
}
\label{fig:fig.e7}
\end{figure*}

\newpage
\begin{figure*}[h!]
\includegraphics[width=0.9\linewidth]{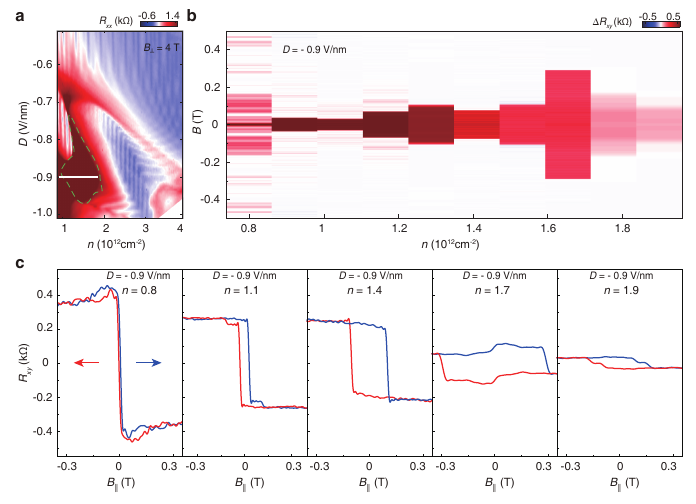}
\caption
{
\textbf{In-plane orbital ferromagnetism at $D$ = - 0.9 V/nm.} 
     \textbf{a}, $R_{xx}$ plotted versus $n$ and $D$ for negative $D$ at $B_\perp$ = 4 T.
     \textbf{b}, Hall resistance difference $\Delta R_{xy}$ as a function of $n$ and $B_{\parallel}$ at $D$ = - 0.9 V/nm, along the white line denoted in \textbf{a}. 
     \textbf{c}, The magnetic-field-dependent $R_{xy}$ at selected carrier density $n$ = 0.8, 1.1, 1.4, 17, 1.9 (in units of $10^{12}$ $\mathrm{cm}^{\mathrm{-2}}$) at $D$ = - 0.9 V/nm , T = 20 mK.
}
\label{fig:fig.e8}
\end{figure*}

\newpage
\begin{figure*}[h!]
\includegraphics[width=1.0\linewidth]{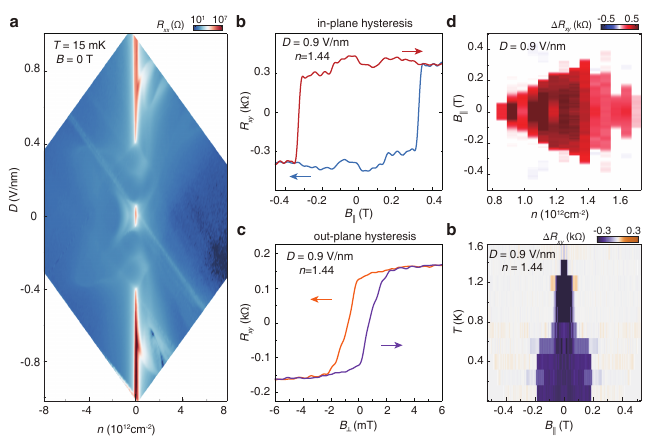}
\caption
{
\textbf{Transdimensional anomalous Hall effect in device B.} 
     \textbf{a}, Map of $R_{xx}$ as a function of $n$ and $D$ at $T$ = 15 mK, and $B$ = 0 T in device B.
     \textbf{b-c}, Antisymmetrized Hall resistance $R_{xy}$ measured as $B_{\parallel}$ (\textbf{b}) and $B_\perp$ (\textbf{c}) is swept back and forth at $n$ = 1.44 x $10^{12}$ $\mathrm{cm}^{\mathrm{-2}}$ and $D$ = 0.9 V/nm. The pronounced hystereses loop with $B_{\parallel}$ and $B_\perp$ indicate the presence of in-plane orbital magnetization and out-of-plane magnetization, thus TDAHE. 
     \textbf{c}, Hall resistance difference $\Delta R_{xy}$ as a function of $n$ and $B_{\parallel}$ at $D$ = 0.9 V/nm.
     \textbf{d}, Hall resistance difference $\Delta R_{xy}$ as a function of $T$ and $B_{\parallel}$ at $n$ = 1.44 x $10^{12}$ $\mathrm{cm}^{\mathrm{-2}}$ and $D$ = 0.9 V/nm 
}
\label{fig:fig.e9}
\end{figure*}

\newpage
\begin{figure*}[h!]
\includegraphics[width=0.9\linewidth]{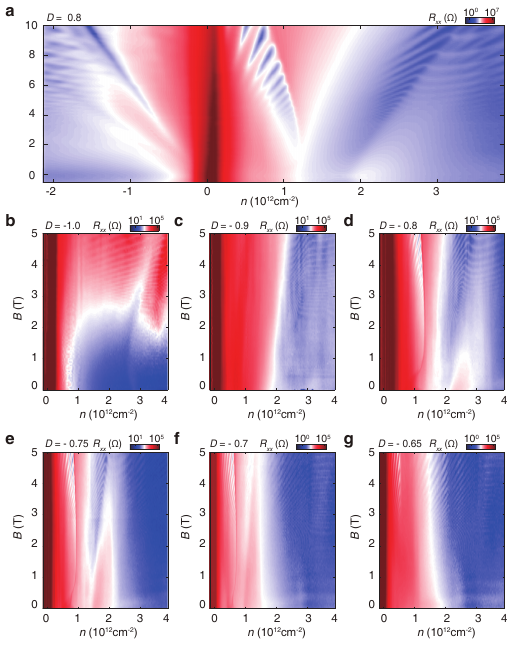}
\caption
{
\textbf{Quantum oscillations near TDAHE regime.} 
     \textbf{a}, Landau fan diagram of $R_{xx}$ at a positive $D$ = 0.8 V/nm.
     \textbf{b-g}, Landau fan diagrams of $R_{xx}$ at negative $D$ values: $D$ = - 1.0 V/nm (\textbf{b}), $D$ = - 0.9 V/nm (\textbf{c}), $D$ = - 0.8 V/nm (\textbf{d}), $D$ = - 0.75 V/nm (\textbf{e}), $D$ = - 0.7 V/nm (\textbf{f}), $D$ = - 0.65 V/nm (\textbf{g}).
}
\label{fig:fig.e10}
\end{figure*}

\newpage
\begin{figure*}[h!]
\includegraphics[width=0.9\linewidth]{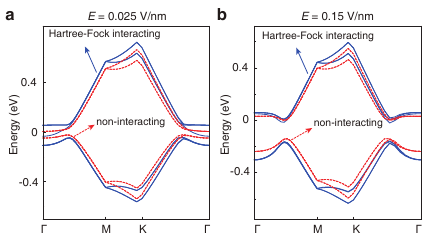}
\caption
{
\textbf{Hartree-Fock calculations of low electric field $E$ and large $E$  band structures.} 
     \textbf{a}, Non-interacting (red dashed curves) and interacting (blue solid curves) band structures at $E$ = 0.025 V/nm, calculated by Hartree-Fock method. 
     \textbf{b}, Non-interacting (red dashed curves) and interacting (blue solid curves) band structures at $E$ = 0.15 V/nm, calculated by Hartree-Fock method. 
}
\label{fig:fig.e11}
\end{figure*}

\newpage
\clearpage

\def\a{\mathbf{a}}
\def\b{\mathbf{b}}
\def\p{\mathbf{p}}
\def\ta{\tilde{\mathbf{a}}}
\def\L{\mathbf{L}}
\def\k{\mathbf{k}}
\def\vr{\mathbf{r}}
\def\G{\mathbf{G}}
\def\Q{\mathbf{Q}}
\def\br{\mathbf{r}}
\def\bd{\mathbf{d}}
\def\kt{\widetilde{\mathbf{k}}}
\def\q{\mathbf{q}}
\def\g{\mathbf{g}}
\def\qt{\widetilde{\mathbf{q}}}
\def\nn{\nonumber \\}
\def\hc{\hat{c}}
\def\hcd{\hat{c}^{\dagger}}
\def\hd{\hat{d}}
\def\hdd{\hat{d}^{\dagger}}
\def\R{\mathbf{R}}

\newcommand{\vect}[1]{\mathbf{#1}}
\newcommand{\ket}[1]{\vert #1 \rangle}
\newcommand{\bra}[1]{\langle #1 \vert}
\newcommand{\bracket}[2]{\langle #1 \vert #2 \rangle}
\newcommand{\eV}{{\text{eV}}}
\newcommand{\eVA}{{\text{eV}\cdot \text{\AA}}}
\newcommand{\bk}{\mathbf{k}}
\newcommand{\bq}{\mathbf{q}}
\newcommand{\btk}{\widetilde{\mathbf{k}}}
\newcommand{\btq}{\widetilde{\mathbf{q}}}
\newcommand{\cop}{\hat{c}}
\newcommand{\dop}{\hat{d}}
\newcommand{\xmark}{\ding{55}}
\def\Red#1{\textcolor{red}{#1}}
\def\Blue#1{\textcolor{blue}{#1}}
\def\I{\uppercase\expandafter{\romannumeral 1}}
\def\II{\uppercase\expandafter{\romannumeral 2}}
\def\III{{\uppercase\expandafter{\romannumeral 3}}}
\def\IV{{\uppercase\expandafter{\romannumeral 4}}}
\def\V{{\uppercase\expandafter{\romannumeral 5}}}
\def\VI{{\uppercase\expandafter{\romannumeral 6}}}
\def\VII{{\uppercase\expandafter{\romannumeral 7}}}
\def\a{\mathbf{a}}
\def\b{\mathbf{b}}
\def\p{\mathbf{p}}
\def\ta{\tilde{\mathbf{a}}}
\def\L{\mathbf{L}}
\def\k{\mathbf{k}}
\def\G{\mathbf{G}}
\def\Q{\mathbf{Q}}
\def\br{\mathbf{r}}
\def\kt{\widetilde{\mathbf{k}}}
\def\q{\mathbf{q}}
\def\g{\mathbf{g}}
\def\qt{\widetilde{\mathbf{q}}}
\def\nn{\nonumber \\}
\def\hc{\hat{c}}
\def\hcd{\hat{c}^{\dagger}}
\def\hd{\hat{d}}
\def\hdd{\hat{d}^{\dagger}}
\def\R{\mathbf{R}}

\centerline{\Large{Supplementary Information for}}
\vspace*{1\baselineskip} 

\centerline{\textbf{\large{Transdimensional anomalous Hall effect in rhombohedral thin graphite}}}

{\centering
  Qingxin Li$^{1}$\textsuperscript,$^{\dagger}$,
  Hua Fan$^{2}$\textsuperscript,$^{\dagger}$,
  Min Li$^{3}$\textsuperscript,$^{\dagger}$,
  Yinghai Xu$^{1}$,
  Junwei Song$^{1}$,
  Kenji Watanabe$^{4}$,
  Takashi Taniguchi$^{5}$,
  Hua Jiang$^{6}$,
  Xin-Cheng Xie$^{6,7,8}$,
  James C. Hone$^{9}$,
  Cory R. Dean$^{10}$,
  Yue Zhao$^{2}$\textsuperscript{$^{\ast}$},
  Jianpeng Liu$^{3,11}$\textsuperscript{$^{\ast}$},
  Lei Wang$^{1,12}$\textsuperscript{$^{\ast}$}
\par}

{\centering
 \textit{$^{1}$National key Laboratory of Solid-State Microstructures, School of Physics, Nanjing University, Nanjing, 210093, China}
 
\textit{ $^{2}$Department of Physics, State key laboratory of quantum functional materials, Guangdong Basic Research Center of Excellence for Quantum Science, Southern University of Science and Technology (SUSTech), Shenzhen 518055, China}
 
 \textit{$^{3}$School of Physical Science and Technology, ShanghaiTech Laboratory for Topological Physics, ShanghaiTech University, Shanghai 201210, China}

 \textit{$^{4}$Research Center for Electronic and Optical Materials, National Institute for Materials Science, 1-1 Namiki, Tsukuba 305-0044, Japan}
 
 \textit{$^{5}$Research Center for Materials Nanoarchitectonics, National Institute for Materials Science, 1-1 Namiki, Tsukuba 305-0044, Japan}
 
 \textit{$^{6}$Interdisciplinary Center for Theoretical Physics and Information Sciences (ICTPIS), Fudan University, Shanghai 200433, China}
 
 \textit{$^{7}$International Center for Quantum Materials, School of Physics, Peking University, Beijing 100871, China}
 
 \textit{$^{8}$Hefei National Laboratory, Hefei 230088, China}
 
 \textit{$^{9}$Department of Mechanical Engineering, Columbia University, New York, NY 10027, USA}
 
 \textit{$^{10}$Department of Physics, Columbia University, New York, NY 10027, USA}

 \textit{$^{11}$Liaoning Academy of Materials, Shenyang 110167, China}
 
 \textit{$^{12}$Jiangsu Physical Science Research Center, Nanjing 210093, China}
\par}

{\centering
 $^{\dagger}$These authors contributed equally to this work.
 $^{\ast}$Corresponding authors, Email: zhaoy@sustech.edu.cn; liujp@shanghaitech.edu.cn; leiwang@nju.edu.cn
 \par}

\maketitle

\setcounter{figure}{0}
\setcounter{equation}{0}

\renewcommand{\theequation}{S\arabic{equation}}

\section{Non-interacting continuum model}

The non-interacting electronic structure of rhombohedral ennealayer graphene can  be described by  a low-energy continuum model Hamiltonian. It is  derived from an atomistic Slater-Koster tight-binding model of multilayer graphene based on carbon's $p_z$-like Wannier orbitals, the Hamiltonian of which is expressed as
\begin{equation}
    H=\sum_{i l\alpha,j l^{\prime}\alpha^{\prime}}
    {-t\left ( 
    \mathbf{R}_i+\bm{\tau}_{\alpha}+l d_0 \mathbf{e}_z
    -\mathbf{R}_j-\bm{\tau}_{\alpha^{\prime}}-l^{\prime} d_0 \mathbf{e}_z 
    \right )} 
    \hat{c}^{\dagger}_{i l \alpha}\hat{c}_{j l^{\prime}\alpha^{\prime}}\;,
\end{equation}
where $i$, $j$ represents lattice sites and $\R_i$, $\R_j$ represents lattice vectors in graphene. $l$ and $l^{\prime}$ are layer indexes while $\alpha$ and $\alpha^{\prime}$ are sublattice indexes. $d_0$ is the interlayer distance and $\mathbf{e}_z$ is an unit vector along out-of-plane direction. $t(\mathbf{d})$ is hopping amplitude between two $p_z$ orbitals displaced by vector $\mathbf{d}$, which is expressed in  the Slater-Kolster form:
\begin{align}
    -t(\mathbf{d}) &= 
    V_{pp\pi}\left [ 1-\left ( \frac{\bd \cdot \mathbf{e}_z}{d} \right )^2  \right ] +V_{pp\sigma}\left ( \frac{\bd \cdot \mathbf{e}_z}{d} \right )^2 \\
    \begin{split}
        V_{pp\pi}&=V_{pp\pi}^0\exp\left ( -\frac{\left | \bd \right |-a/\sqrt{3}}{r_0}  \right ) \\
        V_{pp\sigma}&=V_{pp\sigma}^0\exp\left ( -\frac{\left | \bd \right |-d_0}{r_0}  \right ) 
    \end{split}
\end{align}
where $V_{pp\pi}^0=-2.7\,$eV, $V_{pp\sigma}^0=0.48\,$eV and $r_0=0.184a$ \cite{kpgra}, with $a=2.46\,\text{\AA}$ being graphene's lattice constant. 

Then, we  Fourier transform the Hamiltonian to $\mathbf{k}$ space, and expand the $\mathbf{k}$-space tight-binding Hamiltonian around the  Dirac points of monolayer graphene $\mathbf{K}^{\mu}=-\mu4\pi/3a(1,0)$, where $\mu=\pm 1$ denotes the valley index. Then, we obtain the low-energy continuum model of rhombohedral ennealayer graphene of valley $\mu$ 
\begin{align}
H^{0, \mu}(\mathbf{k})
=    \begin{pmatrix}
h_{\text {intra }}^{0, \mu} & \left(h_{\text {inter }}^{0, \mu}\right)^{\dagger} & 0 & 0&... & 0 \\
h_{\text {inter }}^{0, \mu} & h_{\text {intra }}^{0, \mu} & \left(h_{\text {inter }}^{0, \mu}\right)^{\dagger} & 0&... & 0 \\
0 & h_{\text {inter }}^{0, \mu} & h_{\text {intra }}^{0, \mu} & \left(h_{\text {inter }}^{0, \mu}\right)^{\dagger}  & ... & 0\\
\vdots  & ...& ...& ... & ... & \vdots \\
0 & 0 & 0 & ...& h_{\text {inter }}^{0, \mu} & h_{\text {intra }}^{0, \mu}
    \end{pmatrix}  
\end{align}
We note that for ennealayer rhombohedral graphene, the continuum model $H^{0,\mu}(\mathbf{k})$ is a $18\times 18$ matrix at each wavevector $\mathbf{k}$ including layer and sublattice degrees of freedom. In the above Hamiltonian, $h_{\text {intra }}^{0, \mu}$ is the intralayer part, which is just given by the $\mathbf{k} \cdot \mathbf{p}$ model of the monolayer graphene:
\begin{equation}
h_{\text {intra }}^{0, \mu}=-\hbar v_F^0 \mathbf{k} \cdot \sigma_\mu,    
\end{equation}
where $\hbar v_F^0 \approx 5.253 \mathrm{eV} \text{\AA}$ is the non-interacting Fermi velocity of Dirac fermions extracted from the Slater-Koster tight-binding model \cite{kpgra}, $\mathbf{k}$ is the wave vector expanded around the Dirac point in valley $\mu$, and $\sigma_\mu=\left(\mu \sigma_x, \sigma_y\right)$ is the Pauli matrix defined in sublattice space. The interlayer coupling $h_{\text {inter }}^{0, \mu}$ is expressed as
\begin{equation}
h_{\text {inter }}^{0, \mu}=
\begin{pmatrix}
    \hbar v_{\perp}\left(\mu k_x+i k_y\right) & t_{\perp} \\
\hbar v_{\perp}\left(\mu k_x-i k_y\right) & \hbar v_{\perp}\left(\mu k_x+i k_y\right)\;,
\end{pmatrix}
\end{equation}
where $t_{\perp}=0.34 \mathrm{eV}, \hbar v_{\perp}=0.335 \mathrm{eV} \text{\AA}$ are both extracted from the Slater-Koster hopping parameters described above \cite{kpgra}. In our calculations, we set up a momentum space cutoff (centered at the Dirac point) of $\pi/L_c$ with $L_c=2.25\,$nm, corresponding to an energy cutoff $E_C\sim 1\,$eV for the ennealayer model Hamiltonian given above.

\section{Renormalization of continuum model parameters}

In our work, we mostly focus on the low-energy physics around charge neutrality point of ennealayer graphene. Therefore, we set up a low-energy window marked by $E_C^{*}\sim 0.3\,$eV. Electron-electron interactions are treated using fully unrestricted self-consistent Hartree-Fock approach (see the following section) within this low-energy window within $E_C^*$; while outside $E_C^*$, $e$-$e$ interactions are approximately treated as perturbations to single-particle Hamiltonian outside $E_C^*$. We call the electrons within and outside $E_C^*$ can be partitioned as ``low-energy electrons"  and ``remote-band" electrons. While our primary investigation centers on low-energy phenomena, it is essential to acknowledge that occupied electrons in the remote bands outside $E_C^*$ would significantly influence the low-energy physical properties when electron-electron interactions are considered. The filled electronic states in the remote bands (outside $E_C^*$) mediate their effects through long-range Coulomb potentials, thereby modifying the fundamental characteristics of low-energy electrons through interband correlations. This interaction-driven renormalization process fundamentally alters the effective description of the system. Specifically, when constructing an effective low-energy single-particle Hamiltonian (within the energy cutoff $E_C^*$) incorporating effects from filled remote-band electrons, one finds that its parameters generally acquire enhanced magnitudes compared to those derived from non-interacting models. A paradigmatic example of this enhancement mechanism manifests in graphene's electronic structure: extensive theoretical and experimental studies have demonstrated that the Fermi velocity near the Dirac points becomes substantially renormalized upward due to screening effects from the filled states constituting the Dirac Fermi sea \cite{elias_natphys2011}.

To systematically account for these correlation effects within our theoretical framework, we employ a perturbative renormalization group (RG) methodology. This approach allows us to progressively integrate out high-energy degrees of freedom while preserving the essential physics through scale-dependent parameter renormalization. Rather than reproducing the full technical derivation here, we adopt the established expressions for renormalized low-energy parameters that emerge from this RG treatment, following the comprehensive analysis presented in Refs.~\onlinecite{guo-prb24}.
\begin{subequations}
\begin{align}
 v_F(E_C^*)&=v_F^0 \left(1+\frac{\alpha_0}{4\epsilon_r}\log{\frac{E_C}{E_C^*}} \right)\; \label{eq:H-RG-a}\\
 t_{\perp}(E_C^*)&=t_{\perp}\,\left(1+\frac{\alpha_0}{4\epsilon_r}\log{\frac{E_C}{E_C^*}}\right)\; \label{eq:H-RG-c}\\
 v_{\perp}(E_C^*)&=v_{\perp}\;.
 \label{eq:H-RG-f}
\end{align}
\label{eq:H-RG}
\end{subequations}
Here $E_C\sim 2\,$eV is the largest energy cutoff of the continuum model above which the Dirac-fermion description to graphene would no longer be valid. $\alpha_0=e^2/4\pi\epsilon_0\hbar v_F$ is the effective fine structure constant of graphene, with $\hbar$ being reduced Planck constant and $\epsilon_0$ denoting vacuum permittivity. We refer the readers to Ref.~\onlinecite{guo-prb24} for detailed derivations of the above equations.

\section{Hartree-Fock treatment to electron-electron interactions within the low-energy window}

We consider the Coulomb interactions
\begin{equation}
\hat{V}_\text{ee}=\frac{1}{2}\int d^2 r  d^2 r' \sum _{\sigma, \sigma '} \hat{\psi}_\sigma ^{\dagger}(\br)\hat{\psi}_{\sigma '}^{\dagger}(\br ') V (|\br -\br '|) \hat{\psi}_{\sigma '}(\br ') \hat{\psi}_{\sigma}(\br)
\label{eq:coulomb}
\end{equation}
where $\hat{\psi}_{\sigma}(\br)$  is real-space electron annihilation operator at $\br$ with spin $\sigma$, and $V(|\br-\br'|)$ is the Coulomb potential. After transforming to Wannier function basis, and only including the dominant density-density interaction in Wannier representation, then the Coulomb interaction is simplified to
\begin{align}
\hat{V}_\text{ee}=&\frac{1}{2}\sum _{i i'}\sum _{\alpha \alpha '}\sum _{\sigma \sigma '}\hat{c}^{\dagger}_{i, \sigma \alpha}\hat{c}^{\dagger}_{i', \sigma'  \alpha} V_{i  \alpha, i'  \alpha '}\hat{c}_{i', \sigma ' \alpha '}\hat{c}_{i, \sigma \alpha} \nonumber \\
\approx&\frac{1}{2}\sum _{i  \alpha \neq i' \alpha '}\sum _{\sigma \sigma '}\hat{c}^{\dagger}_{i, \sigma  \alpha} \hat{c}^{\dagger}_{i', \sigma '  \alpha '} V_{i  \alpha,i'  \alpha '}\hat{c}_{i', \sigma ' \alpha '}\hat{c}_{i, \sigma  \alpha} \nonumber \;,
\end{align}
where
\begin{align}
&V_{ i  \alpha  , i'  \alpha'} \nonumber \\
&=\int d^2 r d^2 r'  V (|\br -\br '|) \,\phi ^*_{\alpha} (\mathbf{r}-\mathbf{R}_i-\bm{\tau}_{\alpha})\,\phi_{\alpha} (\mathbf{r}-\mathbf{R}_j- \bm{\tau}_{\alpha}) \phi^*_{ \alpha  '}(\br-\mathbf{R}_i'-\bm{\tau}_{\alpha '})\phi _{ \alpha  '}(\br-\mathbf{R}_i'-\bm{\tau} _{\beta '}) \nonumber \\
&\quad \times \chi ^{\dagger}_\sigma \chi ^{\dagger}_{\sigma '}\chi _{\sigma '}\chi _{\sigma} .
\end{align}
Here $i$, $\alpha$, and $\sigma$ refer to Bravais lattice vectors, layer/sublattice index and spin index, and $\hat{c}_{i,\sigma\alpha}$ ($\hat{c}_{i,\sigma\alpha}^{\dagger}$) is electron annhilation (creation) operator in the corresponding Wannier basis. 
Here we neglect on-site Coulomb interactions, i.e. Hubbard interactions. This is because when the atomic on-site Hubbard interaction $U_0$ ($U_0\sim 1\text{-}5\,$eV) is projected to the low-energy states, it is suppressed by an order of $n\Omega_0$, where $n$ is the carrier density of the system and $\Omega_0$ is graphene's unit-cell area. Therefore, at low carrier density $n\sim 10^{12}\,\text{cm}^{-2}$, the effective atomic Hubbard interaction experienced by the low-density electrons $\sim U_0 n\Omega_0\sim U_0\times 10^{-3}$. On the other hand, the long-range, inter-site Coulomb interaction experienced between low-energy electrons $\sim e^2/4\pi\epsilon_{\rm{BN}}\epsilon_0 L_s$ ($\epsilon_{\rm{BN}}\approx 4$ is the relative dielectric constant of hBN substrate), where $L_s\sim\sqrt{1/\pi n}$ is characteristic inter-electron spacing. When the density $n\sim 10^{12}\,\text{cm}^{-2}$, $L_s\sim 5\,nm$, so that the long-range inter-site Coulomb interaction would be much larger than the atomic Hubbard interaction.  In other words,  given that the electron density is low ($10^{12}$~cm$^{-2}$), i.e., a few electrons per supercell, the chance that two electrons meet at the same atomic site is very low. The Coulomb interactions between two electrons are mostly contributed by the inter-site ones. 

In order to model the screening effects to the $e$-$e$ Coulomb interactions, we consider the double-gate screened Coulomb interaction, the Fourier transform of which is expressed as
\begin{equation}
   V^{\text{dg}}(\mathbf{q})\!=\!e^2\tanh(|\mathbf{q}|d_s)/(\,2 \epsilon_{\textrm{BN}}\varepsilon_0 |\mathbf{q}|\,)
 \label{eq:V_thomasfermi}
\end{equation}
where $\mathbf{q}$ denotes wavevector and $d_s$ is the thickness of hexaongal BN susbtrate, which is set to 40\,nm throughout the calculations.
 
Since we are interested in the low-energy bands, the intersite Coulomb interactions can be divided into the intra-valley term and the inter-valley term. The intervalley term is about two orders of magnitude weaker than the intravalley one at low densities $n\sim 10^{12}\,\text{cm}^{-2}$. Eventually, we keep the dominant intravalley long-range Coulomb interaction , which can be expressed as
\begin{equation}
\hat{V}^{\rm{intra}}=\frac{1}{2S}\sum_{\alpha\alpha '}\sum_{\mu\mu ',\sigma\sigma '}\sum_{\bk \bk ' \bq} V^{\text{dg}}(\bq)\,
\hat{c}^{\dagger}_{\sigma \mu \alpha}(\bk+\bq) \hat{c}^{\dagger}_{\sigma' \mu ' \alpha '}(\bk ' - \bq) \hat{c}_{\sigma ' \mu ' \alpha '}(\bk ')\hat{c}_{\sigma \mu \alpha}(\bk)\;,
\label{eq:h-intra}
\end{equation}
where $S$ is the total area of the entire system, and the operator $\hat{c}_{\sigma\mu\alpha}(\bk)$ ($\hat{c}_{\sigma\mu\alpha}(\bk)^{\dagger}$) annihilates (creates) an electron of valley $\mu$, spin $\sigma$, layer/sublattice $\alpha$ at wavevector $\bk$ relative the Dirac point $\mathbf{K}^{\mu}=-\mu4\pi/3a(1,0)$. 

The electron annihilation operator can be transformed from the original basis to the energy-band basis (Bloch eigenfunction basis):
\begin{equation}
\hat{c}_{\sigma\mu \alpha}(\bk)\equiv  =\sum_n C_{\mu\alpha ,n}(\bk)\,\hat{c}_{\sigma \mu,n}(\bk)\;,
\label{eq:transform}
\end{equation}
where $C_{\mu \alpha,n}(\bk)$ is the wavefunction coefficient of the $n$-th Bloch eigenstate at $\bk$ of valley $\mu$: 
\begin{equation}
\ket{\sigma \mu, n; \bk}=\sum_{ \alpha }C_{\mu  \alpha,n}(\bk)\,\ket{\sigma, \mu,  \alpha; \bk}\;.
\end{equation}
We note that the non-interacting Bloch functions are spin degenerate. Using the transformation given above, the dominant intra-valley Coulomb interaction can be written in the band basis as
\begin{align}
\hat{V}^{\rm{intra}}&=\frac{1}{2 S}\sum _{\btk \btk'\btq}\sum_{\substack{\mu\mu' \\ \sigma\sigma'}}\sum_{\substack{nm\\ n'm'}\in E_C^*} \left(\sum _{\mathbf{Q}}\,V (\bq)\,\Omega^{\mu ,\mu'}_{nm,n'm'}(\bk,\bk',\bq)\right) \nonumber \\
&\times \hat{c}^{\dagger}_{\sigma\mu,n}(\bk+\bq) \hat{c}^{\dagger}_{\sigma'\mu',n'}(\bk'-\bq) \hat{c}_{\sigma'\mu',m'}(\bk') \hat{c}_{\sigma\mu,m}(\bk)
\label{eq:Hintra-band}
\end{align}
where the form factor $\Omega ^{\mu,\mu'}_{nm,n'm'}$ of the Coulomb interaction is expressed as
\begin{equation}
\Omega ^{\mu ,\mu'}_{nm,n'm'}(\bk,\bk',\bq)
=\sum _{\alpha\alpha'}C^*_{\mu \alpha,n}(\btk+\btq) C^*_{\mu'\alpha',n'}(\bk'-\bq)C_{\mu'\alpha',m'}(\bk')C_{\mu \alpha,m}(\bk).
\end{equation}
In Eq.~\eqref{eq:Hintra-band}, the summation notation ``$\sum_{nm,n'm'\in E_c^*}$" means that the sum of the Bloch eigenstates is restricted to the low-energy Bloch states within the energy window $E_C^*\sim 0.3\,$eV. In other words, the Coulomb interactions are projected  to the low-energy states within $E_C^*$.
The effects of interactions between the remote-band electrons and low-energy electrons are treated using perturbative RG approach, which lead to renormalized  single-particle parameters for low-energy electrons within $E_C^*$  as described in the previous section.

Then, we make Hartree-Fock approximation to the band-projected interaction Eq.~\eqref{eq:Hintra-band} such that the two-particle Hamiltonian is decomposed into  Hartree and Fock effective single-particle terms. We perform unrestricted self consistent Hartree-Fock calculations, assuming 32 possible initial symmetry-breaking states characterized by the order parameters $\hat{\tau}_{a}\hat{\sigma}_{b} \hat{s}_{0,z}$, with $a, b=0, x, y,z$, where $\hat{\tau}$, $\hat{\sigma}$ and $\hat{s}$ denote Pauli matrices in the valley, sublattice and spin space, respectively. A ${46\times 46}$ $\mathbf{k}$ mesh is used in the Hartree-Fock calculations, with a fixed momentum space cutoff of $\pi/L_c$ with $L_c=2.25\,$nm, corresponding to an energy cutoff $E_C\sim 1\,$eV.

\end{document}